# Challenges and characterization of a Biological system on Grid by means of the PhyloGrid application

R. Isea[1], E. Montes[2], A.J. Rubio-Montero[2] and R. Mayo[2]

[1] *Fundación IDEA, Hoyo de la Puerta, Valle de Sartenejal, Baruta 1080 (Venezuela)*
*risea@idea.gob.ve*
[2] *CIEMAT, Avda. Complutense, 22 - 28040 Madrid (Spain)*
*{esther.montes,antonio.rubio,rafael.mayo}@ciemat.es*

## Abstract

*In this work we present a new application that is being developed. PhyloGrid is able to perform large-scale phylogenetic calculations as those that have been made for estimating the phylogeny of all the sequences already stored in the public NCBI database. The further analysis has been focused on checking the origin of the HIV-1 disease by means of a huge number of sequences that sum up to 2900 taxa. Such a study has been able to be done by the implementation of a workflow in Taverna.*

## 1. Introduction

The determination of the evolution history of different species is nowadays one of the more exciting challenges that are currently emerging in the computational Biology [1]. In this framework, Phylogeny is able to determine the relationship among the species and, in this way, to understand the influence between hosts and virus [2]. As an example, we can mention the work published in 2007 related to the origin of the oldest stumps in HIV/AIDS (apart from the African one), which were found in Haiti [3], so it is clear that the test of vaccines for this disease must include both African and Haitian sequences. This work was performed with MrBayes tool, the bayesian inference method used in PhyloGrid.

With respect to the computational aspect, the main characteristic of the phylogenetic executions is that they are extremely intensive, even more when the number of sequences increases, so it is crucial to develop efficient tools for obtaining optimized solutions. Thus, 1000 taxa (sequences) generate $2.5 \cdot 10^{1167}$ trees, so among all of them the potential consensus tree is found. That is why it is clear to understand the computational challenge that this work represents by studying the origin of the HIV-1 since it is dealing with 2900 sequences.

Several techniques for estimating phylogenetic trees have been developed such as the distance based method (which generates a dendrogram, i.e. they do not show the evolution, but the changes among the sequences), the maximum parsimony technique, the maximum likelihood method and the bayesian analysis (mainly Markov Chain Monte Carlo inference). The latest is based on probabilistic techniques for evaluating the topology of the phylogenetic trees; so it is important to point out that the Maximum Parsimony methods are not statistically consistent, this is, the consensus tree cannot be found with the higher probability because of the long branch attraction effect [4].

In this work, we have worked with MrBayes software [5] for obtaining the phylogenetic trees. It is important to indicate that MrBayes is relatively new in the construction of these trees as the reader can check in the pioneering work of Rannala and Yang in 1996 [6]. This methodology works with the Bayesian statistics previously proposed by Felsentein in 1968 as indicated Huelsenbeck [7], a technique for maximizing the subsequent probability. The reason for using this kind of approach is that it deals with higher computational speed methods so the possible values for the generated trees can all be taken into account not being any of them ruling the others. With respect to the consensus tree found in our work, the value of which was obtained with this methodology, known as parametric bootstrap percentages or Bayesian posterior probability values, we can mention that it gives information about the reproducibility of the different part of the trees, but this kind of data does not represent a significant statistical value [8].

Thus, based on MrBayes tool, the PhyloGrid application aims to offer to the scientific community an easy interface for calculating phylogenetic trees by means of a workflow with Taverna [9]. In this way, the user is able to define the parameters for doing the Bayesian calculation, to determine the model of evolution and to check the accuracy of the results in the intermediate stages. In addition to this, no knowledge from his/her side about the computational procedure is required. More details about this workflow can be found in [10].

As a consequence of this development, several biological results have been achieved in the Duffy domain of Malaria [10] or in the HPV classification [11]. In this work, we study the dependence of a successful determination for a biological system with the number of sequences to be aligned. To do so, the HIV case has been selected taking into account over a thousand different sequences.

## 2. Tools

The structure of the implementation of the different tools present in the PhyloGrid application can be seen in Figure 1. We here briefly explain them.

### 2.1. MrBayes

Bayesian inference is a powerful method which is implemented in the program MrBayes [12] for estimating phylogenetic trees that are based on the posterior probability distribution of the trees. Currently is included in different scientific software suites for cluster computing such as Rocks Cluster (Bio Roll) as well as in other Linux distributions where the user doesn't need to compile it, for example Ubuntu, Gentoo, Mandriva and so on. This easiness allows the program to be ported to the Grid. On the other hand, its main drawback is that its execution for millions of iterations (generations) requires a large amount of computational time and memory. As an example, we can cite that in the work from Cadotte et al. [13] 143 angiosperms were determined by means of MrBayes. In that work, four independent Markov Chains were run, each with 3 heated chains for 100 million generations, the authors sampled the runs every 10,000 generations and used a burning of 70 million steps to generate a majority rule consensus tree that was used to calculate PDC. Such a methodology is basically the same for other calculations and the number of iterations will depend on the convergence from the chains. In fact, this kind of scientific calculation is extremely complex from a mathematical point of view, but also from the computational one where the phylogenetic estimation for a medium size dataset (50 sequences, 300 nucleotides for each sequence) typically requires a simulation for 250,000 generations, which normally runs for 50 hours on a PC with an Intel Pentium4 2.8 GHz processor.

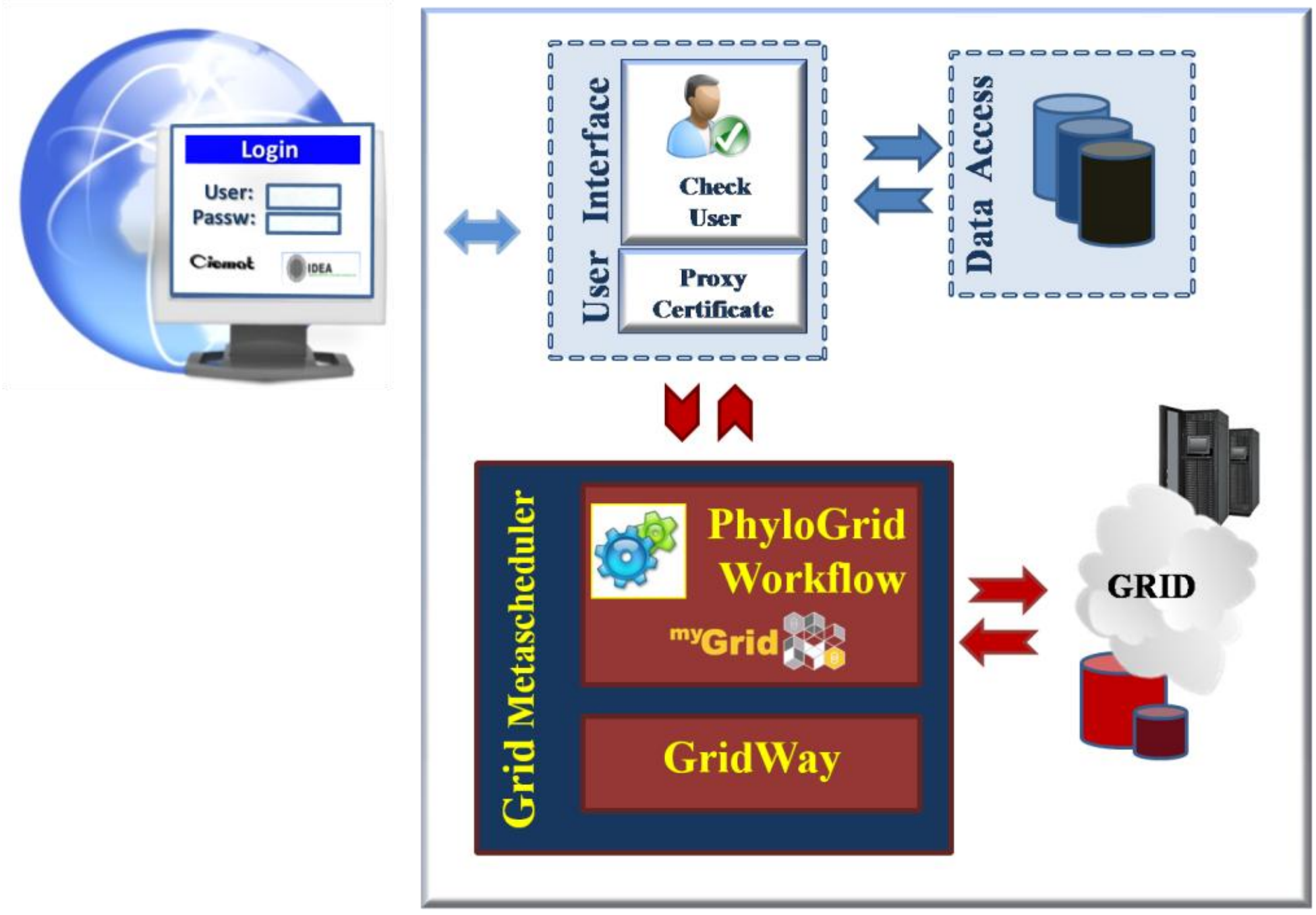

**Figure 1. EELA Schema of PhyloGrid**

Even more, the recent publication by Pérez et al. [14] shows that the study of 626 domains of prokaryotic genomes, due to they are a major class of regulatory proteins in eukaryotes, represents by itself a computational challenge that must be limited to the analysis of the results instead of trying to delimit the number of obtained results by the number of sequences that implies the computational time.

## 2.2. Gridsphere

The Gridsphere project aims to develop a standard based portlet framework for building web portals and a set of portlet web applications. It is based on Grid computing and integrates into the GridSphere portal framework a collection of gridportlets provided as an add-on module, i.e. a portfolio of tools that forms a cohesive "grid portal" end-user environment for managing users and provides access to information services. Grid security based on public key infrastructure (PKI) and emerging IETF and OASIS standards are also well-defined characteristics. In this way, it is important to mention that the users are able to access the Grid by means of an implementation of the JSR 168 portlet API standard. As a consequence, they can also interact with other standards such as those of GT4 and all of its capabilities.

An important key in this application is that the researcher who use it can do it with his/her personal Grid user certificate ("myproxy" initialization), the execution of which is already integrated in the Gridsphere release. There is also the possibility of running the jobs with a proxy directly managed by the Administrator that would be renewed from time to time in order to allow longer jobs to be ended. Thus, all the technical details are transparent for the user, so all the methodology is automated and the application can either be run directly by a certified user or letting Gridsphere to assign him a provisional proxy, registered in a map log.

Within the portal framework and in a future release, the possibility for doing a multiple alignment of the sequences will be available for the user.

### 2.3. Taverna, a tool for the implementation of Workflows

The workflow is fully built in Taverna [9] and structured in different services that are equivalent to the different sections that are run in a common MrBayes job and performs a complete calculation just building the input file by means of the construction of a common Grid `jdl` file. The front end to the final user is a web portal built upon the Gridsphere framework. This solution makes very easy to the researcher the calculations of molecular phylogenies by no typing at all any kind of command.

The main profit that these kind of workflows offer is that they integrate in a common development several modules (tools) connected one to each other. Even more, the deployment of the application with Taverna allows any researcher to download the complete workflow, making easy in this way their spread in the scientific community.

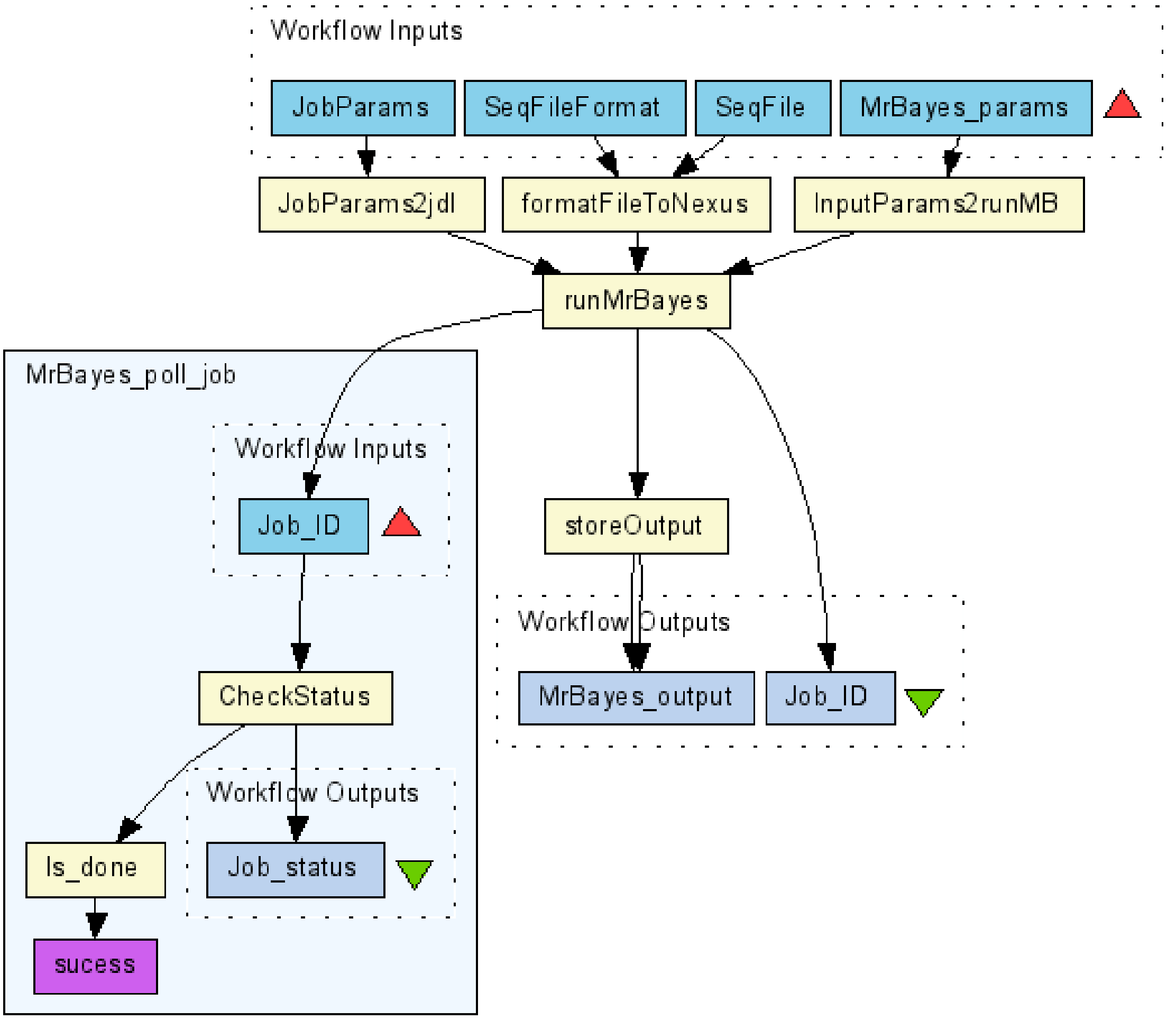

**Figure 2. The Taverna Workflow used in this work**

This is so because Taverna allows the user to construct complex analysis workflows from components located on both remote and local machines, run these workflows on their own data and visualise the results. To support this core functionality it also allows various operations on the components themselves such as discovery and description and the selection of personalised libraries of components previously discovered to be useful to a particular application. Finally, we can indicate that Taverna is based on a workbench window where a main section of an Advanced Model Explorer shows a tabular view of the entities within the workflow. This includes all processors, workflow inputs and outputs, data connections and coordination links. For using Taverna, it is necessary to create a web service for the required application that will be integrated into the software. Lately, this web service will call MrBayes inside the workflow.

## 2.4. The PhyloGrid Workflow: a short description

The structure of the Taverna workflow can be seen in Figure 2, the structure of which has been improved since the first works performed with the PhyloGrid application [10]. Once the user has logged into the PhyloGrid portal, registered a valid proxy and introduced all the data needed for his/her job, a background process performs the execution of the PhyloGrid workflow with the input provided by the user.

The workflow receives several inputs: the MrBayes parameters (labelled as `MrBayes_params` in Fig. 2) that define the evolutionary model that will be used in the analysis and the settings for Markov Chain Monte Carlo (MCMC) analysis; the parameters needed to construct the appropriate Grid job file (`JobParams`); the input file with the aligned sequences (`SeqFile`) to analyse; and, the format of the file (`SeqFileFormat`).

The first three processors of the workflow (`JobParams2jdl`, `formatFileToNexus` and `InputParams2runMB`) perform some tasks prior to MrBayes execution. Thus, as aforementioned, the processor named `JobParams2jdl` creates the appropriate file for the job submission to Grid; the `formtFileToNexus` processor, if necessary, converts the file with the aligned sequences to NEXUS format (not available in the first PhyloGrid release); and the `InputParams2runMB` constructs an executable file with MrBayes commands from the `MrBayes_params`.

The output of these processors is sent to the core processor of the workflow, i.e. `runMrBayes`. This processor submits a MrBayes analysis job (see below the Methodology section). A call to a nested workflow (`MrBayes_poll_job`) is included to check for job status, and wait if job is not ended. When the MrBayes analysis job is finished, the `runMrBayes` processor receives a notification from the `MrByes_poll_job` processor.

As a final step, an additional processor is included to store the output of the MrBayes analysis in a Storage Element. This step is needed due to the large size that MrBayes output files can reach. Once the output files are stored in a SE, the workflows execution ends and the user can retrieve the results from the PhyloGrid portal.

## 2.5. GridWay: The next step

Even when PhyloGrid is producing reliable scientific results, the improvement of the tool is a key factor for this work team. That is why we are planning to introduce GridWay [15], the standard metascheduler in Globus Toolkit 4, into the whole structure of the application in order to perform the submission of jobs in a more efficient way. Since GridWay enables large-scale, reliable and efficient sharing of heterogeneous computing resources managed by different LRM systems, it is able to determine in real time the nodes that best fit the characteristics of the submitted jobs. In the case of MrBayes and its parallel mode, this functionality is not only limited to match in any resource the adequate tags into the information systems whatever the installed mpich version the node had, but also to select the best resources by means of several statistical methods based on accounting of previous executions. As a consequence, the calculation will be done in the lowest possible time.

At the same time, GridWay can recover the information in case of a cancelled job since it creates checkpointers. It also migrates automatically failed jobs to other Grid resources. Thus, it performs a reliable and unattended execution of jobs transparently to Taverna that simplifies the workflow development and increases the global performance of PhyloGrid.

The porting process of GridWay into PhyloGrid will be done by means of the plugins already available for Globus in Taverna that allow the use of resources by a standardized way.

## 3. Methodology

Once the user has determined his/her work area and has connected to the Internet Network, new PhyloGrid jobs can be submitted. For doing so, he/she logs in the application and a new window is then open. In this page, the user is able to define the name of the job to be submitted as well as its description, to upload the file with the alignment, to select the model of evolution and the number of iterations with a fixed frequency and, finally, to run the experiment.

The Taverna workflow builts the script that will rule the process with the input data once they are set in the MrBayes format. In addition, as parameters the user must configure: the selected model for the evolution of the system (labelled as the `lset` part/command in the example below); the number of simultaneous, completely independent, analyses (`nruns` set to 1); the number of generations that will be performed in the calculation (`ITER`); and, finally, the location of the file where the sequences are present (`file-aligned.nex`). To the date, this file must be in Nexus format, but in further releases the workflow will be able to translate to a NEXUS format the input alignment if it is written in any other kind of format (Clustal, Phylip, MSA and so on). Thus, our example would be written as:

```
begin mrbayes;
set autoclose=yes nowarn=yes;
execute /Path-to-file/file-aligned.nex;
lset nst=6 rates=gamma;
mcmcp nruns=1 ngen=ITER samplefreq=FREQ;
mcmc ;
mcmc ;
mcmc ;
end;
```

This is, the workflow must perform its load section by section. Since the first two instructions are always the same for any kind of calculation, the workflow has to begin with the third one (`execute`...) making a call for bringing the input data, i.e. the aligned sequences to be studied, and following with the rest of commands.

PhyloGrid needs to know the evolution model, so the fourth instruction (`lset`...) is used. Here, two options are possible: to allow the researcher to select a specific one or to allow the workflow to do so. The fifth instruction (`mcmcp nruns…`) sets the parameters of the MCMC analysis without actually starting the chain. In our example, it sets the number of independent analyses (`nruns`) and the number of executions to be made (the `ITER` data, one million of iterations for example) with a concrete frequency (`FREQ`). This command is identical in all respects to the `mcmc` one, except that the analysis will not start after this command is issued. The following instructions (`mcmc`) perform the MCMC analysis (three consecutive ones in our example), which will be able also to be monitored. All these options are able to be changed by the final user, who at the beginning of the process has simply defined the evolutionary model that will be used in the analysis, the settings for MCMC analysis and the name of the input file with the aligned sequences in NEXUS format. By default, if the name of the output files is not provided, MrBayes sets the corresponding extensions for them (files that will be generated adding the `.p`, `.t` and `.mcmc` extensions to the name of the input file).

The job must always have the final command `end` since it is mandatory for MrBayes.

Once the workflow has started, MrBayes automatically validates the number of iterations meanwhile it begins to write the output file and sets the burning value for generating the phylogenetic trees. When the whole calculation is ended and the packed output files are stored in an appropriate Storage Element, it can be downloaded by the user for further analysis.

## 4. Results and conclusions

PhyloGrid has been used to calculate the Phylogeny of the HIV-1 sequences stored in the NCBI database, the number of which is up to 2900. All these sequences were previously aligned with MPIClustal. The calculation lasted for 576 hours on 20 cores inside 5 Intel Xeon X5365 (3GHz, 4 MB L2 cache per 2 cores in the quad-core processor) with 2GB of memory per core. The output file had 835 MB, the tree of which can be seen in Figure 3.

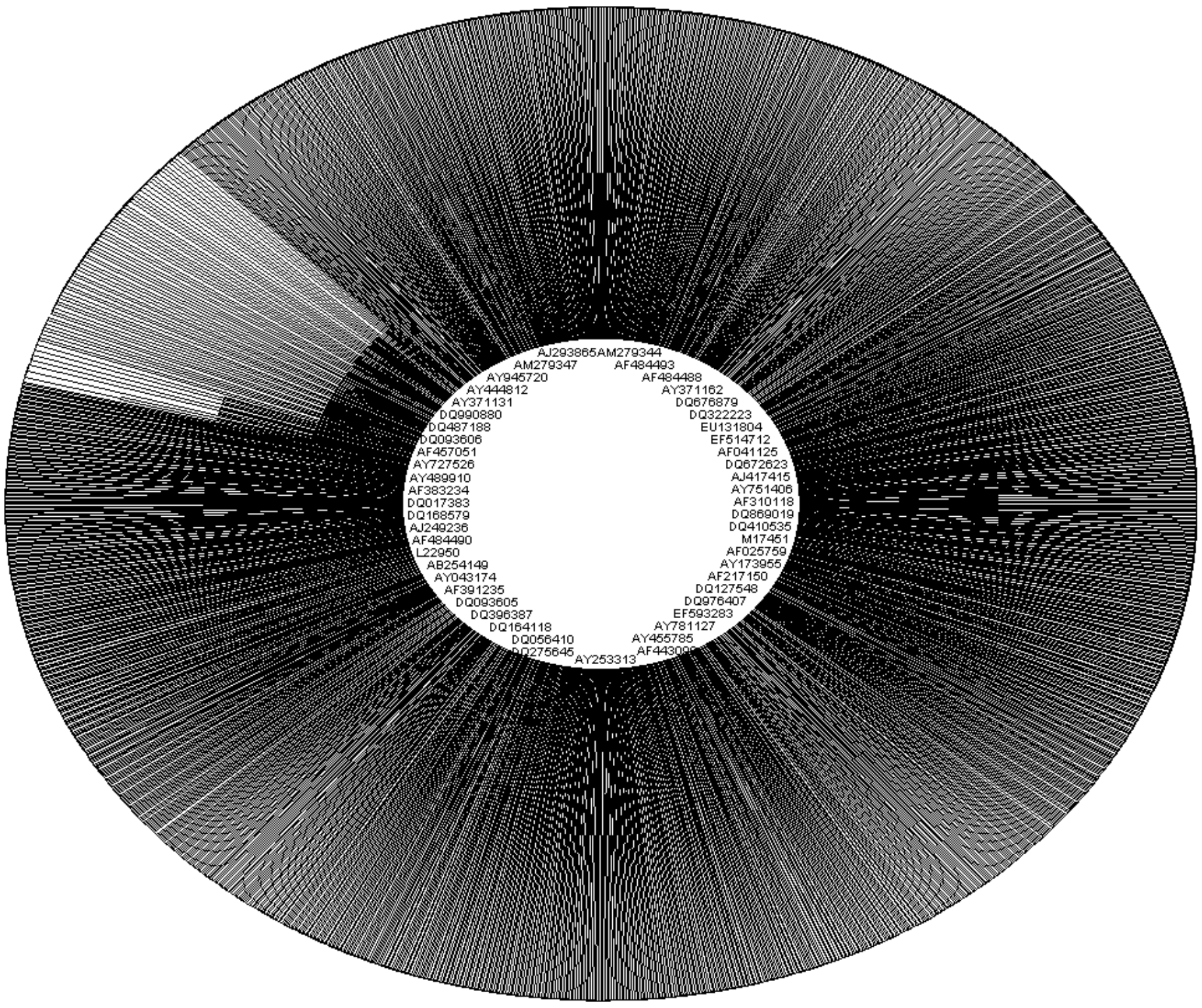

**Figure 3. Phylogenetic tree construction using PhyloGrid of HIV-1 sequences**

To the knowledge of the authors, these results are the first that are totally performed for the HIV counting on its whole number of sequences. In a future, the results will be analysed separately, but we can point out the periodicity of the branchs in the obtained trees, so a further study on the obtained patterns will be done in order to simplify the analysis of the results.

As can be seen for the parameters of the performed calculation (output file size or time consumed), it is clear that Grid is a useful paradigm that allows the Biologists to cope with such a complex research.

## Acknowledgements

This work makes use of results produced by the EELA-2 project (http://www.eu-eela.eu), co-funded by the European Commission within its Seventh Framework Programme and the VII Cuba-Venezuela Scientific Collaboration Framework and the Annual Operative Plan of Fundación IDEA.